\documentclass[twocolumn,showpacs,prl]{revtex4}
\usepackage{graphicx}
\usepackage{dcolumn}
\usepackage{bm}

\begin{document}

\title{Improving noise threshold for optical quantum computing \\
with the EPR photon source}
\author{Z.-H. Wei$^{1,2}$, Y.-J. Han$^{2}$, C. H. Oh$^{1}$, and L.-M. Duan$^{2}$}
\affiliation{$^{1}$Centre for Quantum Technologies, National
University of Singapore, Singapore 117542\\$^{2}$Department of
Physics and MCTP, University of Michigan, Ann Arbor, Michigan 48109}

\begin{abstract}
We show that the noise threshold for optical quantum computing can
be significantly improved by using the EPR-type of photon source.
In this implementation, the detector efficiency $\eta _{d}$ is
required to be larger than $50\%$, and the source efficiency $\eta
_{s}$ can be an arbitrarily small positive number. This threshold
compares favorably with the implementation using the single-photon
source, where one requires the combined efficiency $\eta _{d}\eta
_{s}>2/3$. We discuss several physical setups for realization of
the required EPR photon source, including the photon emitter from
a single-atom cavity.
\end{abstract}

\maketitle

Optical quantum computing has raised significant interest in
recent years \cite{a1,1,1b,2,3,4,5,6,7,8}, in particular after the
innovative proposal by Knill,Laflamme, Milburn (KLM), who show
that the feed-forward from high-efficiency photon detectors
provides the required nonlinearity for the optical gate operations
\cite{1}. The architecture of the gate in the original KLM\
proposal is somewhat complicated and the required detection
efficiency is very high for scalable computation \cite{1,2}. This
requirement gets significantly relaxed with an improved approach
to optical quantum computing \cite{3,4}, based on the
cluster-state model for quantum computation \cite{5}. The
threshold inefficiency for the photon detection (or in general for
the photon loss errors) is improved to the $1\%$ level with this
cluster state approach, as estimated in Ref. \cite{6}. The next
step of significant improvement has been made recently with the
proposal of a clever architecture of tree graphs for efficient
correction of the dominant photon loss errors \cite{7}. In this
approach, it is shown that the photon loss, as measured by the
source efficiency $\eta _{s}$ and the detection efficiency $\eta
_{d}$, only need to fulfil the threshold requirement $\eta
_{s}\eta _{d}>2/3$. The photon loss in the memory or during the
optical manipulation can be taken into account by combining their
effects with the detection efficiency, which reduces the value of
the effective efficiency $\eta _{d}$.

In this paper, we improve the noise threshold for photon loss in optical
quantum computation with a less stringent requirement of $\eta _{s}>0$ and $%
\eta _{d}>1/2$. Furthermore, we eliminate the challenging requirement of the
number-resolving photon detectors assumed in the previous work \cite{7}. The
significant improvement in this paper is achieved with a simple change in
the implementation: we use the EPR\ photon source instead of the
single-photon source as usually assumed in optical quantum computing. We
then discuss several physical setups for generating the required EPR\ photon
source.

In the cluster-state approach to optical quantum computation, the
central task is to create a large-scale graph state that is
universal for quantum computation (the single-bit gates are
considered to be easy and can be implemented with simple linear
optical elements with a high accuracy) \cite {5}. If we require
the computation to be inherently robust to the photon loss errors,
the underlying graphs for the graph states need to have special
architecture as shown in Ref. \cite{7}. These graph states can be
generated efficiently through some simple linear-optics quantum
gates \cite{8,9}. In particular, the type-II fusion gates are
robust to the photon loss errors \cite{8}. Although these gates
are probabilistic in nature, they can lead to efficient buildup of
arbitrary graph states \cite{7,9b}.

The type-II\ fusion gate eats two photons for each application of the gate
(the photons are absorbed by the two detectors). To connect two graph
states, each of $n$ qubits, with the type-II fusion gate, the output graph
state has the qubit number $2n-2$. To have the qubit number increasing with
application of the gates, one needs to have $n\geq 3$. Therefore, one needs
to start with graph states initially having three photons, which are just
the three-photon GHZ states. Although the three-photon GHZ correlation has
been demonstrated before in the coincidence basis \cite{10}, the states
there can not be used for optical quantum computation as they only survive
in the post-selected Hilbert space which lead to problem in the scaling. For
optical quantum computation, a critical requirement is to realize \textit{%
free }three-photon GHZ states with the vacuum component as small as possible.

Ref. \cite{7} has shown how to generate the independently degraded (ID) GHZ
state from the single-photon source described by the density operator $\rho
_s=(1-\eta _s)|vac\rangle \langle vac|+\eta _s|1\rangle \langle 1|,$ where $%
|vac\rangle $ denotes the vacuum component and $\eta _s$ is the
source efficiency. The ''ID'' state is degraded from the perfect GHZ
state with each photon in the state subject to independent loss with
the same loss rate $f$. The generated ''ID'' GHZ state has an
effective loss rate $f=1-\eta _s/\left( 2-\eta _s\eta_d\right) $
\cite{7}. The ''ID'' states can be connected with the type-II
fusion, yielding larger graph state with the same effective
loss rate $f$. This loss rate $f$, combined with the detection efficiency $%
\eta _d$ for the final single-bit measurements, need to fulfil the threshold
requirement $\left( 1-f\right) \eta _d>1/2$, which leads to $\eta _s\eta
_d>2/3$.

\begin{figure}[tbh]
\centerline{\includegraphics[angle=0,width=3.0in]{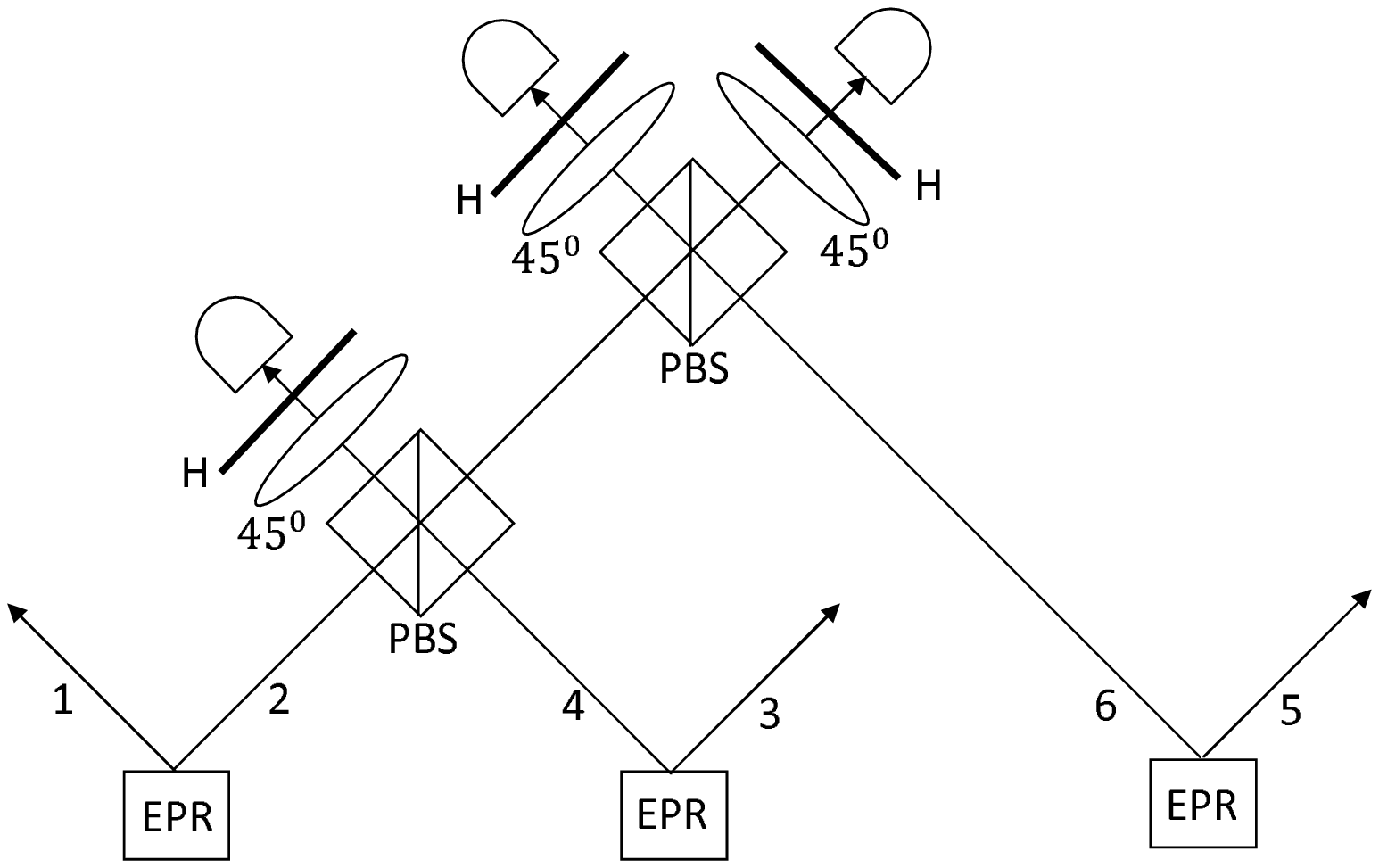}}
\caption{The construction of a free three-photon GHZ state based on
the EPR photon source. The input modes 1 and 2, 3 and 4, 5 and 6,
are in an imperfect EPR state with vacuum components. The photons in
the modes 2, 4, and 6, first go through polarization beam splitters
(PBS), $45^o$-degree polarization rotators, and horizontal (H)
polarizers, and then are detected by single photon detectors. If
each detector registers a photon, the modes 1, 3 , and 5 are
projected onto the GHZ state.}
\end{figure}

Here, instead of the single-photon source, we start with an
imperfect EPR\ state with the source efficiency $\eta _{s}$,
described by the density operator,
\begin{equation}
\rho _{EPR}=(1-\eta _s)|vac\rangle \langle vac|+\eta _s|EPR\rangle \langle
EPR|
\end{equation}
where $|EPR\rangle =$ $(|H_1H_2\rangle +|V_1V_2\rangle )/\sqrt{2}$ denotes
the standard EPR state. We can generate a three-photon GHZ state
\begin{equation}
|GHZ\rangle _{135}=1/\sqrt{2}(|H_1H_3H_5\rangle +|V_1V_3V_5\rangle )
\end{equation}
with the setup shown in Fig. 1, using three pairs of the imperfect
EPR\ state $\rho _{EPR}$. The process is probabilistic and it
succeeds if the three detectors each register a horizontally
polarized photon. In this case, we need to have at least one photon
coming from each input state $\rho _{EPR}$, so the vacuum component
in $\rho _{EPR}$ only influences the success probability, and has no
contribution to the final state when the process succeeds. The
generated GHZ state has no vacuum component (and thus no photon loss
with the above loss rate $f=0$), and these GHZ states can be used to
build up large-scale graph states with the type-II fusion gates. So
the threshold requirement now
is given by $\eta _d>1/2$, which is independent of the source efficiency $%
\eta _s$ in the initial state $\rho _{EPR}$. We only require $\eta _s>0$, so
that the preparation of the GHZ\ states succeeds with a finite probability
given by $P_s=\eta _s^3\eta _d^3/32$ (note that the success probability is $%
\eta _s^3\eta _d^3/256$ for the case of single-photon source). The
finite success probability for preparation of the GHZ\ state does
not affect the scaling and only leads to a constant overhead for
overall quantum computation. Notice also that in the setup shown
in Fig. 1, the photon detectors do not need to resolve the photon
numbers, as no more than one photon can hit each detector in the
event of ''success''. This is different from the case of
single-photon source, where more challenging number-resolving
photon detectors need to be assumed.

Now we discuss several physical setups for possible implementation
of the EPR\ photon source described in Eq. (1). The photon pairs
generated from the spontaneous parametric down conversion (SPDC)
are usually written in the form of Eq. (1) \cite{10,11}. However,
there is an important point that needs to be clarified. For the
photon pairs generated from the SPDC, there is a small probability
to get two (or more) EPR pairs. Although this double EPR
probability is small, it leads to a serious problem. The density
operator for the photon pairs from the SPDC can be written in the
form
\begin{eqnarray}
\rho _{s} =(1-\eta _{s})|vac\rangle \langle vac|+\eta _{s}|EPR\rangle
\langle EPR|  \nonumber \\
+(x\eta _{s}^{2}/2)(|EPR\rangle \langle EPR|)^{\otimes 2}+...,
\end{eqnarray}
where for a Possionian distribution $x=1$ (which is typically the case for
the SPDC). If we input three of this type of states to the setup shown in
Fig.1, after detection on the modes 2, 4, 6, we can analyze the output state
from the modes 1, 3, 5. We assume the source efficiency $\eta _{s}\ll 1$. In
this case, up to the order of $\eta _{s}^{3}$ (any orders lower than this
can not give the three counts on the detectors 2, 4, 6), the following terms
can make a contribution to the registered photon counts: (i) $|EPR\rangle
,|EPR\rangle ,|EPR\rangle $ (one EPR\ pair from each of three inputs); (ii) $%
|vac\rangle ,|EPR\rangle ,|EPR\rangle ^{\otimes 2}$ and its
permutations (one input is in the vacuum whereas another input has
double EPR pairs). So, conditional on a photon count registered on
each of the three detectors, the output state for the three modes
1,3,5 is given by (unnormalized)
\begin{widetext}
\begin{eqnarray}
\rho _{out}=&|GHZ\rangle _{135}\langle GHZ|+(\frac{x}{2}(1-\eta
_s))(|H_1V_1H_5\rangle \langle H_1V_1H_5|+|H_1V_1V_3\rangle \langle
H_1V_1V_3|\nonumber+|H_3V_3H_1\rangle \langle
H_3V_3H_1|\\
&+|H_3V_3V_5\rangle \langle H_3V_3V_5|+|H_5V_5V_1\rangle \langle
H_5V_5V_1|+|H_5V_5H_3\rangle \langle H_5V_5H_3|)+O(\eta _s).
\end{eqnarray}
\end{widetext}

The terms proportional to $x$ in this equation comes from the contribution
of the case (ii), and the last term $O(\eta _s),$ which is negligible when $%
\eta _s$ is small, comes from the higher order contributions
(remember the photon detectors are not number resolving). The
state is not a ID GHZ\ state, and the terms proportional to $x$
have no photon in some mode while two photons in the other mode.
These terms, after a series of type-II diffusion gates, lead to
complicated error patterns for the final graph state, which is
difficult to correct with the photon detectors. So the state can
not be used for optical quantum computation unless $x$ is small,
which requires sub-Possionian distribution in the input state
$\rho _s$ in Eq. (3). For the conventional SPDC, unfortunately it
has Possionian distribution with $x=1$. One possibility, which
could lead to the state in Eq. (3) with a sub-Possionian
distribution, is using the dipole blockade in an atomic ensemble
\cite{12}. By laser driving an atomic ensemble, one can get an
EPR\ photon source in the form of Eq. (3), very similar to the
process in the SPDC \cite{13,14}. Normally, one also has $x=1$ in
the atomic ensemble, however, if we make use of the Rydberg levels
in the atoms, one can excite only one EPR pair (the double
excitations can be suppressed by the dipole blockade resulting
from the strong dipole-dipole interaction between the Rydberg
atoms \cite{12}), and the resulting output state will have the $x$
terms suppressed.

To completely suppress the double EPR terms, an experimentally
easier method is to use a single dipole in an optical cavity. One
can trap a single atom or a single ion in a high finesse cavity,
or grow a single quantum dot in a semiconductor cavity
\cite{16a,16}. The EPR state of the photons can be generated by
laser exciting a single atom from the side of the cavity
\cite{16}. An example configuration is shown in Fig. 2
\cite{15,16}, and the photon EPR\ state has been generated from a
single trapped atom in a recent experiment based on this type of
configuration \cite{16}. In this example, we assume one of the
hyperfine states of the atoms has the hyperfine spin $F=1$ (this
is the case, for instance, for the $^{87}$Rb or $^{23}$Na atom and
the Yb$^{+}$ ion). The atom is initially prepared in the level
$\left|
F=1,m=0\right\rangle $. A laser pulse from the side of the cavity with $\pi $%
-polarization drives the vertical transition, and pushes the atom to the
excited state (say, a $P$ state with the hyperfine level $\left|
F=2,m=0\right\rangle $). The atom emits a photon to the cavity mode, either $%
\sigma ^{+}$\ or $\sigma ^{-}$ polarized, and decays back to the the
corresponding ground state $\left| F=1,m=-1\right\rangle $ or $\left|
F=1,m=1\right\rangle $. A laser pulse with $\pi $-polarization again drives
the vertical transition, and pushes the atom to the excited state $\left|
F=1,m=-1\right\rangle $ or $\left| F=1,m=1\right\rangle .$ The atom then
decays back to the $\left| F=1,m=0\right\rangle $ level, emitting another $%
\sigma ^{-}$\ or $\sigma ^{+}$ polarized photon. The two photons
emerging from the decay of the cavity modes are naturally in an EPR
state
\begin{eqnarray}
|EPR\rangle _{12}=&(|\sigma _{1}^{+},\sigma _{2}^{-}\rangle +|\sigma
_{1}^{-},\sigma _{2}^{+}\rangle )/\sqrt{2}\nonumber\\
=&(|H_{1}V_{2}\rangle +|V_{1}H_{2}\rangle )/\sqrt{2},
\end{eqnarray}
where $|H\rangle $ and $|V\rangle $ are equal linear superposition of $%
|\sigma ^{+}\rangle $ and $|\sigma ^{-}\rangle $ with different sign. If we
take into account the finite laser excitation probability and decay of the
atom to other atomic levels, the photon source is described by $\rho _{EPR}$
in Eq. (1) with a finite source efficiency , and we can tolerate a large
amount of error due to this finite efficiency as we explained before. For
this setup, we have the double excitation probability $x=0$, as with a short
pulse, a single atom can emit only a single photon.

\begin{figure}[tbh]
\centerline{\includegraphics[angle=0,width=3.0in]{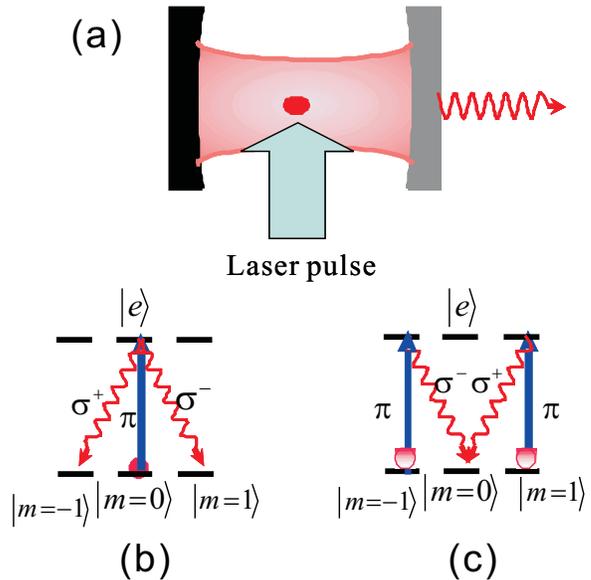}}
\caption{An illustration of generation of the EPR photon source from a
single atom (ion) trapped in a cavity. (a): The schematic setup. (b) and
(c): The excitation configuration for the first and the second laser pulses,
which produce two photons entangled in the polarization basis.}
\end{figure}

In this setup, another type of photon loss (such as the photon
absorption or scattering by the cavity mirror) could lead to missing
of only one photon. The state in general should be represented by
the density operator
\begin{eqnarray}
\rho_c=&p_0|vac\rangle \langle vac|+\frac{p_1}{2}(|H_1\rangle
\langle
H_1|+|V_1\rangle \langle V_1|) \nonumber \\
&+\frac{p_2}{2}(|H_2\rangle \langle H_2|+|V_2\rangle \langle
V_2|)+p_3|EPR\rangle \langle EPR|.
\end{eqnarray}
Without the terms $p_{1}$ and $p_{2}$, the state $\rho _{c}$
reduces to the standard state $\rho _{EPR}$ with the source
efficiency $\eta _{s}=p_{3}$. The terms $p_{1}$ and $p_{2}$
represent the possibilities that only one of the photons in the
EPR pair is lost, and $p_{1}$ and $p_{2}$ in general are not
equal. For instance, in the above scheme, after the first
excitation by the laser pulse, if the atom decays to a different
hyperfine level through spontaneous emission, we will not get the
first photon from the cavity output and in this case there is also
no second photon (as the second driving pulse can not excite the
atom any more due to the large off resonance given by the
hyperfine splitting). However, if the atom only decays to a
different hyperfine level after the second laser excitation, we
have no
second photon but still have the first photon. So in general, we have $%
p_{1}>p_{2}$. In the GHZ\ preparation scheme shown in Fig. 1, we
detect the mode 2 (which has a larger photon loss) and output the
mode 1. When the input EPR pairs are represented by the general
state $\rho _{c}$ in Eq. (6), we can derive the output state for the
modes 1, 3, and 5 after detection on the modes 2, 4, and 6 in Fig.
1. After some tedious but straightforward calculations, we find that
the final output state is given by
\begin{widetext}
\begin{eqnarray}
\rho _{out}^{c} =&(1-f)^{3}|GHZ\rangle \langle GHZ|+\frac{f(1-f)^{2}}{2}%
(|H_{1}H_{3}\rangle \langle H_{1}H_{3}|+|V_{1}V_{3}\rangle \langle
V_{1}V_{3}|+|H_{1}H_{5}\rangle \langle
H_{1}H_{5}|\nonumber\\&+|V_{1}V_{5}\rangle \langle
V_{1}V_{5}|+|H_{5}H_{3}\rangle \langle H_{5}H_{3}|
+|V_{5}V_{3}\rangle \langle
V_{5}V_{3}|)+\frac{f^{2}(1-f)}{2}(|H_{1}\rangle \langle
H_{1}|\nonumber\\&+|V_{1}\rangle \langle V_{1}|+|H_{3}\rangle
\langle H_{3}|+|V_{3}\rangle \langle V_{3}|+|H_{5}\rangle \langle
H_{5}|+|V_{5}\rangle \langle V_{5}|)+f^{3}|vac\rangle \langle vac|.
\end{eqnarray}
\end{widetext}
This is exactly an independently degraded (ID) GHZ state with the loss
probability $f=\frac{p_2}{p_2+p_3}$. This ID-GHZ states can be used to
construct large scale graph states with the same loss probability by
applying the type-II fusion gate. So the threshold requirement becomes $%
\left( 1-f\right) \eta _d>1/2$. If we take the detector efficiency
$\eta _d=75\%$, the ratio $p_2/p_3$ is required to be
$p_2/p_3<1/2$ (note that the vacuum component $p_0$ can be
arbitrarily large). It is pretty routine to achieve such a
requirement with the state of the art cavity technology \cite
{16a,16}.

In summary, we have shown that with use of an imperfect EPR\
photon source, the threshold on photon loss for optical quantum
computation can be significantly improved and we can eliminate the
requirement of using the number-resolving photon detectors. We
discuss physical setups where the required EPR photon source can
be implemented. In particular, the single-dipole cavity provides a
clean EPR source with no double excitations, and the requirements
are pretty realistic with the state-of-the-art cavity technology.

This work was supported by the ARO through MURI, the grant from
the Centre for Quantum Technology, the WBS grant under contract
no. R-710-000-008-271, the AFOSR\ through MURI, the DARPA, and the
IARPA.

\end{document}